# NESTED HEAD-TAIL VLASOV SOLVER

A. Burov, FNAL, Batavia, IL 60510, USA

*Abstract*

Nested Head-Tail (NHT) is a Vlasov solver for transverse oscillations in multi-bunch beams. It takes into account azimuthal, radial, coupled-bunch and beam-beam degrees of freedom affected by arbitrary dipole wakes, feedback damper, beam-beam effects and Landau damping.

## INTRODUCTION

Collective instabilities impose a serious limitation on beam intensity in circular accelerators, storage rings and colliders. That is why quantitative description of the instabilities is required for understanding of both existing and future machines. Especially dangerous are transverse instabilities, normally resulting in a beam loss. In principle, there are two approaches for their modelling. The older approach is based on the Vlasov equation, a dynamic equation for a phase space density of collisionless plasma; the equation was suggested in 1938. In fact, this equation was first written not for plasma, but for gravitating cosmic objects by J. Jeans in his publications of years 1913-1915, see a historical article of M. Henon [1]. So far, solutions of the Vlasov equation were limited by rather simple cases (see e. g. Ref [2]), insufficient to tell about complicated reality of multi-bunch beams with feedbacks, beam-beam effects and octupoles. That is why the second approach to beam stability problems, macroparticle simulations, attracted more and more attention, driven by continuing burst of computational powers. For colliders, such codes as HEADTAIL [3] and BeamBeam3D [4] are known and used for more than 10 years. Attractiveness of the macroparticle tracking programs is related to their similarity to real beams; they appear to be as close to reality as possible, allowing rather straightforward introduction of all the factors influencing beam stability. However, an attempt of these direct simulations of reality has its drawback: convergence typically requires a big number of macroparticles per bunch, at the order ~$10^6$ or so. For thousands bunches per beam in the collider, it makes a required study of multi-dimensional space of parameters so far impossible by means of the macroparticle tracking – even with the help of parallel computations by powerful supercomputers.

This limitation of the tracking methods brings us back to the Vlasov equation with a motivation to develop more sophisticated methods of its solution, where all important factors of reality would be properly taken into account. Nested Head-Tail (NHT) suggests an attempt in this direction [5]. Its name points to its primary idea: solutions of the Vlasov equation are sought as expansions over conventional head-tail functions, defined at a set of nested rings in the longitudinal phase space. Since the impedance, feedback, and coherent beam-beam are described by linear response functions, their analysis is reduced to a standard eigensystem problem that is solved instantaneously, as soon as the related matrices are defined. After this is done, growth rates and tune shifts of all potentially dangerous modes are known. However, with any feedback, that pure linear system is still unstable: its stabilization requires some Landau damping, an intrinsic self-stabilization mechanism caused by nonlinearity of single-particle motion. In general, these anharmonicities lead to very complicated equations (see e. g. Eq.(6.179) in Ref. [2]). However, for many practical cases, the anharmonicities may be treated as perturbations of the linear system. When Landau damping moves the coherent tune shifts not by much, it can be found as a perturbation. That is how Landau damping is treated in this paper, allowing finding thresholds of the instabilities, with both octupoles and beam-beam nonlinearities taken into account. Fortunately, for many practical cases this perturbative approach to Landau damping is justified. For pure transverse nonlinearities it leads to well-known dispersion relations, see Ref. [6] and references therein. Otherwise, more general form of the dispersion relation has to be applied, see a section "Landau Damping". Numerous examples of NHT applications for LHC are shown in the last two sections.

## BASIS FUNCTIONS

Bunch longitudinal distribution inside a linear potential well can be represented by a sequence of concentric rings, or air-bags, as it was suggested by Ref. [7]. It appears to be optimal to keep all the rings equally populated, so the phase space density is reflected by variable distances between them. To do that, the phase space has to be divided onto certain number of equally populated areas; then a weighted-average radius has to be found for each area. The number of rings has to reflect the wake properties. Unless the wake is extremely microwave, the default NHT choice of $n_r = 5$ shown in Fig. 1 should work reasonably well. On these concentric rings, a conventional set of head-tail harmonics $\psi_{l\alpha}$ is defined as a basis for phase space density perturbations (see e.g. Ref. [2], Eq.(6.175)):

$$\psi_{l\alpha} = \exp(il\phi + i\chi_\alpha \cos\phi - iQ_x\omega_0 t)\,;$$
$$\chi_\alpha = Q'_x \omega_0 \tau_\alpha / \eta\ . \qquad (1)$$

Here $\phi$ is the synchrotron phase, $l$ is azimuthal head-tail harmonic number, $1 \le \alpha \le n_r$ is radial harmonic number, $\omega_0$ is angular revolution frequency, $Q_x = \omega_x / \omega_0$ is the lattice betatron tune, $\chi_\alpha$ stays for

* FNAL is operated by Fermi Research Alliance, LLC under Contract No. De-AC02-07CH11359 with the United States Department of Energy.

the set of the head-tail phases for the rings with radii $\tau_\alpha$ in time units, $Q'_x$ is the chromaticity, $c$ is speed of light, $\eta = \gamma_t^{-2} - \gamma^{-2}$ is the slippage factor.

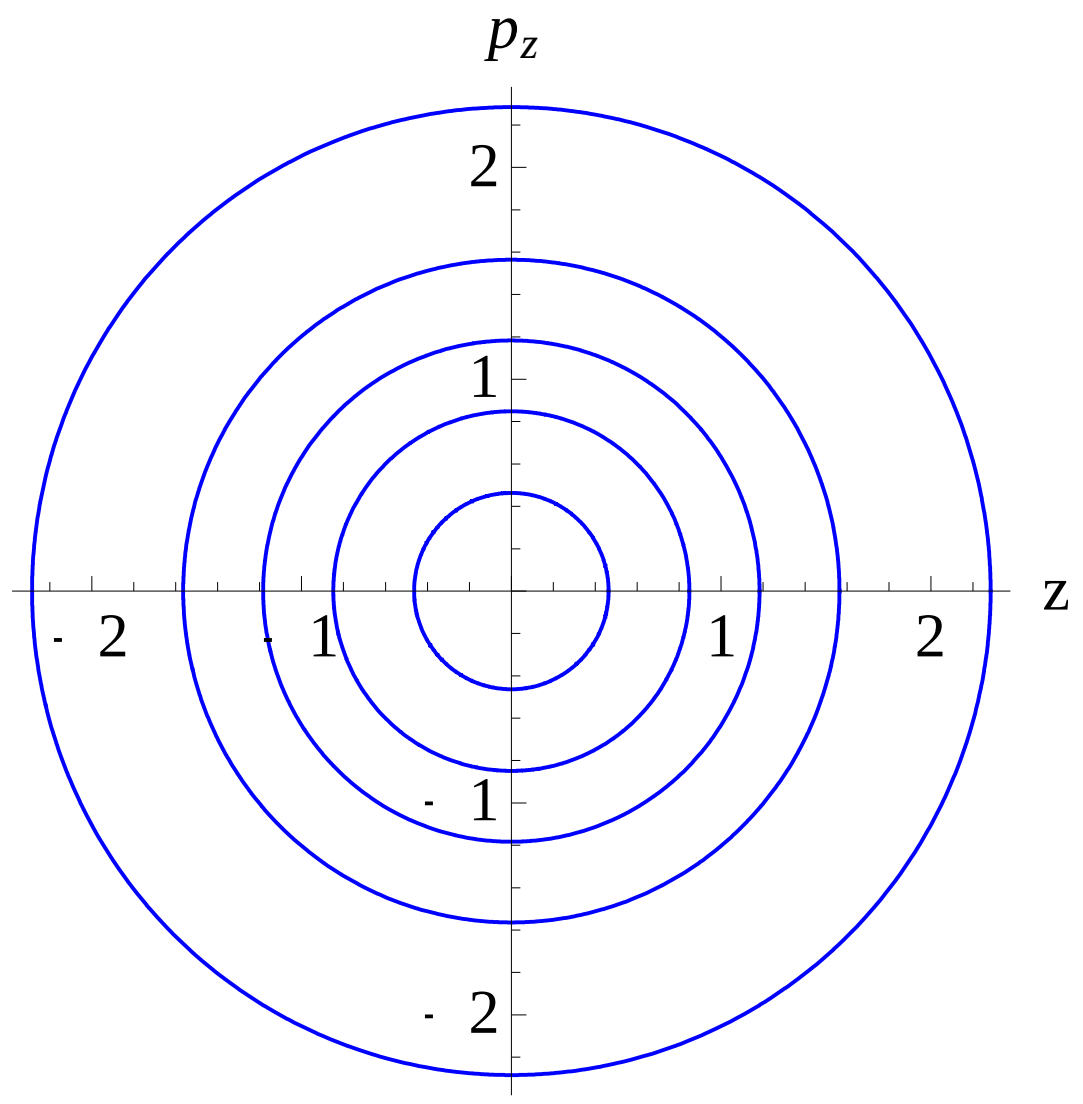

Figure 1: Bunch longitudinal Gaussian distribution inside a linear potential well presented by equally populated nested air-bags; the coordinate $z$ and momentum $p_z$ are measured in their rms units. Note that the radii are not equidistant.

## SINGLE BUNCH EIGENSYSTEM

Any dipole perturbation of the bunch distribution can be expanded over this nested head-tail basis. Following Ref. [2], assuming time dependence $\propto \exp(-i\Omega t - iQ_x\omega_0 t)$, Eq. (6.183) of that reference can be presented as:

$$q\mathbf{X} = \hat{\mathbf{S}}\cdot\mathbf{X} - i\hat{\mathbf{Z}}\cdot\mathbf{X} - ig\hat{\mathbf{F}}\cdot\mathbf{X} \qquad (2)$$

Here, $\mathbf{X}$ is the perturbation vector with components along the basis functions $\psi_{l\alpha}$, matrix $\hat{\mathbf{S}}$ reflects harmonic oscillations inside the RF potential well, matrix $\hat{\mathbf{Z}}$ reflects a single-turn transverse impedance, $g$ is a damper gain, and $\hat{\mathbf{F}}$ is a flat-wake matrix:

$$\begin{aligned}
&\hat{S}_{lm\alpha\beta} = l\delta_{lm}\delta_{\alpha\beta}\ ;\\
&\hat{Z}_{lm\alpha\beta} = i^{l-m}\frac{\kappa}{n_r}\int_{-\infty}^{\infty} d\omega Z(\omega) J_l(\omega\tau_\alpha - \chi_\alpha) J_m(\omega\tau_\beta - \chi_\beta)\ ;\\
&\kappa = \frac{N_b r_0 R_0}{8\pi^2\gamma Q_x Q_s}\ ;\\
&\hat{F}_{lm\alpha\beta} = \frac{i^{m-l}}{n_r} J_l(\chi_\alpha) J_m(\chi_\beta)\ .
\end{aligned} \qquad (3)$$

Here $Z(\omega)$ is the transverse impedance, $N_b$ is the number of particles per bunch, $r_0$ is the classical radius, $R_0 = c/\omega_0$, $Q_s = \omega_s/\omega_0$ is the synchrotron tune. The dimensionless eigenvalue $q = \Omega/\omega_s$ is the coherent tune shift in units of the synchrotron tune. The flat-wake matrix $\hat{\mathbf{F}}$ describes any dynamic response whose variation over the bunch length can be neglected [8]. Note that flat wakes do not satisfy causality condition of conventional wakes, since they are not related to an instantaneous action of the same-beam particles. Examples of flat wakes are those of dampers whose bandwidth is much smaller than an inverse bunch length, inter-bunch and multi-turn wakes, beam-beam responses for beta-function exceeding the bunch length.

Taken with the damper gain $g$, the term $g\hat{\mathbf{F}}$ describes the response of a flat damper, which sees only centroids and whose kicks are bunch-flat. The term "flat" as it is applied here to the damper assumes space or time flatness, so it is not to be confused with the idea of a high bandwidth damper, whose response is flat in a frequency domain. To avoid this confusion, the term "flat" is applied below for the time or space domain only. Fourier-image of a flat wake is delta-function, so the expression for the flat-wake matrix $\hat{\mathbf{F}}$ in Eq. (3) follows from the impedance matrix $\hat{\mathbf{Z}}$ after a substitution $Z(\omega) \propto \delta(\omega)$, defining the gain $g$ as a damping rate in units of the synchrotron frequency. In case the damper is so broadband that its response is not flat over the bunch length, its actual non-flat wake function has to be used, making its matrix different from $\hat{\mathbf{F}}$.

## MULTI-BUNCH EIGENSYSTEM

Inter-bunch wake functions $W(s)$ are normally flat thanks to suppression of high-frequency cavity modes. If so, inter-bunch terms in the equation of motion can be described by means of the flat-wake matrix $\hat{\mathbf{F}}$, used in Eq. (3) for the flat damper. Inter-bunch wake fields can be conventionally summarized for equidistant bunches whose oscillation amplitudes differ only by a phase factor $\exp(2\pi i\mu k/M)$, where $0 \le k \le M-1$ is a bunch number, $0 \le \mu \le M-1$ is the inter-bunch mode number, and $M$ is the number of bunches. Inter-bunch interaction contributes its own term to the originally single-bunch Eq. (2):

$$\begin{aligned}
&q\mathbf{X} = \hat{\mathbf{S}}\cdot\mathbf{X} - i\hat{\mathbf{Z}}\cdot\mathbf{X} - ig\hat{\mathbf{F}}\cdot\mathbf{X} + \hat{\mathbf{C}}\cdot\mathbf{X}\ ;\\
&\hat{\mathbf{C}} = 2\pi\kappa\tilde{W}_\mu\hat{\mathbf{F}}\ ;\\
&\tilde{W}_\mu = \sum_{k=1}^{\infty} W(-ks_0)\exp(ik\varphi_\mu)\ ;\\
&\varphi_\mu = 2\pi(\mu + Q_x)/M\ ;\quad s_0 = 2\pi R_0/M\ .
\end{aligned} \qquad (4)$$

These equations reduce the multi-bunch problem to a set of single-bunch ones since every inter-bunch mode $\mu$ is treated separately by Eq. (4), where the sought-for eigenvector **X** has the same structure as for the single-bunch problem of Eq. (2).

The multi-bunch problem of Eq. (4) assumes that the damping rate $g$ is identical for all the inter-bunch modes, being the same as for the single-bunch case; in other words, it assumes that the damper is a bunch-by-bunch one. In case it is not so, the damping rate becomes a function of the inter-bunch mode number $\mu$. This function can be found in a way similar to the wake inter-bunch coefficients $\tilde{W}_\mu$. In the time domain, the damper response can be described by means of its wake function $G(s)$, associated with the gain frequency profile $G_\omega$ by means of the Fourier transform:

$$G(s) = \int_0^\infty G_\omega \cos(\omega s / c) d\omega / \pi \,, \tag{5}$$

assuming this wake function to be even (remember that flat wakes are not causal). With that wake, the inter-bunch mode coefficients $\tilde{G}_\mu$ analogous to the wake coefficients $\tilde{W}_\mu$ can be found:

$$\tilde{G}_\mu = G(0) + 2\sum_{k=1}^{\infty} G(ks_0)\cos(k\varphi_\mu) \tag{6}$$

In Eq. (4), the mode-independent gain $g$ has to be changed on a mode-dependent one, $g \to g_\mu$, and the latter can be expressed as

$$g_\mu = g_0 \tilde{G}_\mu / \tilde{G}_0 \,, \tag{7}$$

where $g_0$ stays for the damping rate of the inter-bunch mode $\mu = 0$.

In case the bunches are not all equidistant, similar summation of the inter-bunch wakes can be done using the "train theorem" [8]. This procedure requires two steps. First, an eigensystem $\{p, \mathbf{R}\}$ of the total inter-bunch wake (sum of the conventional and damper wakes) has to found:

$$p_\mu \mathbf{R}_\mu = (2\pi\kappa\hat{\mathbf{W}} - ig\hat{\mathbf{G}})\mathbf{R}_\mu . \tag{8}$$

Second, for the examined inter-bunch mode, its eigenvalue $p_\mu$ has to be used in Eq. (4) by means of a substitution

$$2\pi\kappa\tilde{W}_\mu - ig_\mu \to p_\mu . \tag{9}$$

According to the "damper theorem" [8], for a sufficiently high damper gain, the inter-bunch (and beam-beam) collective interaction can be neglected, thus reducing the multi-bunch problems (4), (10), (13), (14) to the single-bunch case (2), provided the inter-bunch and beam-beam wakes are flat.

## BEAM-BEAM RESPONSE

For colliders, there is one more source which complicates beam dynamics: beam-beam effects. There are two different aspects in the beam-beam interaction: coherent ("strong-strong") and incoherent ("weak-strong") ones. The coherent beam-beam effect is associated with a collective response of one beam to another. This response works as a specific wake function, coupling the two beams. In case the bunch length is much smaller than the beta-function in the collision point, this interaction can be treated as flat in the sense of this paper. Flat kicks of oncoming bunches are constant over the bunch length, being determined by centroid offsets and being independent of all other details of intra-bunch oscillations. The incoherent aspect of beam-beam interactions causes additional anharmonicities of single-particle motion, thus contributing to Landau damping. In this section, only the coherent part is considered, while the incoherent one is left for a chapter below where all optics nonlinearities are treated together in a framework of the dispersion relation.

Inclusion of the coherent beam-beam interaction doubles dimension of the problem. Each beam can be described by the same Eq. (4), where an additional beam-beam coupling term has to be added in the right-hand side.

A simplest case of beam-beam response is one of a single flat collision in a single interaction region (IR). In this situation, Eq. (4) is modified as

$$\begin{aligned} q\mathbf{X}_1 &= \hat{\mathbf{S}}\cdot\mathbf{X}_1 - i\hat{\mathbf{Z}}\cdot\mathbf{X}_1 - ig_\mu\hat{\mathbf{F}}\cdot\mathbf{X}_1 + \hat{\mathbf{C}}\cdot\mathbf{X}_1 + \\ &\quad \xi\mathbf{X}_1 - \xi\hat{\mathbf{F}}\cdot\mathbf{X}_2 \;; \\ q\mathbf{X}_2 &= \hat{\mathbf{S}}\cdot\mathbf{X}_2 - i\hat{\mathbf{Z}}\cdot\mathbf{X}_2 - ig_\mu\hat{\mathbf{F}}\cdot\mathbf{X}_2 + \hat{\mathbf{C}}\cdot\mathbf{X}_2 + \\ &\quad \xi\mathbf{X}_2 - \xi\hat{\mathbf{F}}\cdot\mathbf{X}_1 \;. \end{aligned} \tag{10}$$

Here $\mathbf{X}_{1,2}$ are perturbation vectors for the two beams, and $\xi$ is a beam-beam tune shift. The first beam-beam term in the right-hand side, $\xi\mathbf{X}_{1,2}$, describes a linear part of the incoherent tune shift; it does not actually play a role, being identical to an external quadrupole, equal for both beams. The second beam-beam term, $-\xi\hat{\mathbf{F}}\cdot\mathbf{X}_{2,1}$, describes the beam-beam wake under assumption of it flatness. For long-range (parasitic) collisions with an impact parameter $\rho$ and local beta-function $\beta_x$

$$\xi = \frac{N_b r_0 \beta_x}{2\pi\gamma\rho^2 Q_s} = \frac{N_b r_0}{2\pi\tilde{\rho}^2 \varepsilon_n Q_s} \,, \tag{11}$$

where $\varepsilon_n$ is a normalized rms emittance and $\tilde{\rho} = \rho\sqrt{\gamma / \varepsilon_n \beta_x}$ is a normalized bunch separation; sign corresponds to proton-proton collisions in the same $x$-plane.

For the head-on collisions of round beams

$$\xi = -\frac{N_b r_0}{8\pi\varepsilon_n Q_s} \,; \tag{12}$$

for non-round beams see Ref. [9]. A factor of $Q_s$ enters in Eqs. (11), (12) due to a convention of this paper to measure all tune shifts in units of the synchrotron tune.

For more than one collision per IR, the beam-beam term has to be modified, taking into account that the inter-bunch phases of the oncoming beam are not identical; they vary according to the considered coupled-bunch mode $\mu$. Note that for equidistant bunches, beam-beam collisions do not break the inter-bunch mode structure $\propto \exp(2\pi i \mu k / M)$ since that is just a consequence of the translational invariance. Summation over $2K+1$ mirror-symmetric kicks results in a following modification of Eq. (10):

$$
\begin{aligned}
q\mathbf{X}_1 &= \hat{\mathbf{S}}\cdot\mathbf{X}_1 - i\hat{\mathbf{Z}}\cdot\mathbf{X}_1 - ig_\mu\hat{\mathbf{F}}\cdot\mathbf{X}_1 + \hat{\mathbf{C}}\cdot\mathbf{X}_1 + \\
&\quad \xi\mathbf{X}_1 - \xi\,\mathrm{K}_\mu\hat{\mathbf{F}}\cdot\mathbf{X}_2\ ; \\
q\mathbf{X}_2 &= \hat{\mathbf{S}}\cdot\mathbf{X}_2 - i\hat{\mathbf{Z}}\cdot\mathbf{X}_2 - ig_\mu\hat{\mathbf{F}}\cdot\mathbf{X}_2 + \hat{\mathbf{C}}\cdot\mathbf{X}_2 + \\
&\quad \xi\mathbf{X}_2 - \xi\,\mathrm{K}_\mu\hat{\mathbf{F}}\cdot\mathbf{X}_1\ ; \\
\mathrm{K}_\mu &= \sum_{k=-K}^{K}\tilde{\rho}_k^{-2}\cos(2\pi\mu k/M) \Big/ \sum_{k=-K}^{K}\tilde{\rho}_k^{-2}\ ,
\end{aligned}
\tag{13}
$$

where $\xi$ is the total linear part of the incoherent beam-beam tune shift.

One more complication of the beam-beam coupling appears when there is more than one IR. To avoid unnecessarily cumbersome expressions, let's assume that there are only two IRs. In case the IRs are completely identical, and the betatron phase advances between them are equal for the two beams, the coupling terms of the IRs simply add together in Eq. (13). If one of these identical IRs is tilted by 90° relatively another, both the total incoherent and the total coherent beam-beam terms are cancelled, provided the phase advance between them of the beam one is equal to that of the beam two. If the phase advances are not equal, the incoherent terms vanish, but coherent do not. The last situation was realized at the LHC, with its orthogonal beam crossing planes for the two main IRs and significant difference between the inter-IR phase advances of the beam one and beam two, $\psi = \psi_1 - \psi_2$ [10, 11]. Thus, for the LHC case, the dynamic equations (13) have to be modified to

$$
\begin{aligned}
q\mathbf{X}_1 &= \hat{\mathbf{S}}\cdot\mathbf{X}_1 - i\hat{\mathbf{Z}}\cdot\mathbf{X}_1 - ig_\mu\hat{\mathbf{F}}\cdot\mathbf{X}_1 + \hat{\mathbf{C}}\cdot\mathbf{X}_1 - \\
&\quad \xi\, b\mathrm{K}_\mu\hat{\mathbf{F}}\cdot\mathbf{X}_2\ ; \\
q\mathbf{X}_2 &= \hat{\mathbf{S}}\cdot\mathbf{X}_2 - i\hat{\mathbf{Z}}\cdot\mathbf{X}_2 - ig_\mu\hat{\mathbf{F}}\cdot\mathbf{X}_2 + \hat{\mathbf{C}}\cdot\mathbf{X}_2 - \\
&\quad \xi\, b^*\mathrm{K}_\mu\hat{\mathbf{F}}\cdot\mathbf{X}_1\ ; \\
b &= 1 - \exp(i\psi)\ .
\end{aligned}
\tag{14}
$$

These equations yield an eigensystem for transverse dipole multi-bunch oscillations of one or two coupled beams for the given wakes/impedances, gain frequency profile and the collision scheme. Eigenvalues $q$ give the total coherent tune shifts resulted from a combined action of the single- and multi-bunch wakes, damper and beam-beam response.

For the LHC, the Nested Head-Tail program is usually run with 5 radial rings, 21 azimuthal head-tail basis functions, Eq.(1), and 15 representative inter-bunch modes, yielding $5\cdot 21\cdot 15 = 1575$ collective modes for a single beam and twice more for two coupled ones. After computation of the single-bunch impedance matrix $\hat{\mathbf{Z}}$, which needs to be done only once for the given impedance model, the solution of the eigensystem problem takes about 1 second with Wolfram Mathematica 7 or higher installed on an average laptop.

## LANDAU DAMPING

The coherent tune shifts $q$ are found above under a condition of pure harmonic oscillations for all the three degrees of freedom. In general, anharmonicities significantly change mode structure, driving it far from the harmonic case. If so, the considered harmonic solution appears to be useless, and the problem has to be solved from scratch. However, if the anharmonicities are not that large, they can be treated as perturbations to the eigenvalues $q$; the latter can be used as zero approximation. At first approximation, a tune shift driven by the perturbation can be conventionally expressed as a diagonal matrix element of the perturbation term with the unperturbed basis. If the absolute value of this tune perturbation is small enough, the perturbation approach is justified.

For arbitrary non-degenerate matrices, this problem was solved in Ref. [12]. This solution can be expressed as follows. Let $\hat{\mathbf{A}}$ be a non-degenerate matrix, and $\mathbf{V}, \mathbf{U}$ be eigenvectors of this matrix and its Hermit conjugation:

$$\hat{\mathbf{A}}\cdot\mathbf{V} = \hat{\lambda}\cdot\mathbf{V};\quad \hat{\mathbf{A}}^\dagger\cdot\mathbf{U} = \hat{\lambda}^*\cdot\mathbf{U} \tag{15}$$

With all different eigenvalues $\hat{\lambda}$, the two sets of vectors are dual (biorthogonal):

$$\mathbf{U}_j^\dagger\cdot\mathbf{V}_k = \delta_{jk}\mathbf{U}_k^\dagger\cdot\mathbf{V}_k\,. \tag{16}$$

Let $\tilde{\mathbf{A}}$ be a matrix, close to $\hat{\mathbf{A}}$, their difference is proportional to a small parameter. Then, at first order by this parameter, $k$-th eigenvalue $\tilde{\lambda}_k$ of the matrix $\tilde{\mathbf{A}}$ satisfies the following equation:

$$\mathbf{U}_k^\dagger\cdot(\tilde{\mathbf{A}} - \tilde{\lambda}_k\hat{\mathbf{I}})\cdot\mathbf{V}_k = 0\,, \tag{17}$$

where $\hat{\mathbf{I}}$ is the identity matrix. In other words, at the first order of the small parameter, the eigenvalues of the perturbed matrix $\tilde{\mathbf{A}}$ can be found from state-averaging of the perturbed equations with the unperturbed eigenvectors. If the unperturbed matrix $\hat{\mathbf{A}}$ is Hermitian, then $\mathbf{V} = \mathbf{U}$; otherwise the two basis sets are different. This algebraic result suggests treating of incoherent nonlinearities as perturbations, giving rise to the question about its justification.

Nonlinear terms yield an incoherent frequency spread causing Landau damping. When non-linearity is high enough, the beam is stable. The instability threshold corresponds to a case when the growth rate computed for a purely linear situation is equal to the Landau damping rate. Thus, at the threshold, the nonlinear perturbation of the coherent tune is equal to its imaginary part. That is why justification of the perturbation method for the threshold computation requires the imaginary part of the coherent tune shift *q* to be small compared to its real part. Assuming this is true, the method of perturbation is justified for threshold computation.

Let $\mathbf{x}$ be an NHT perturbation vector of a small fraction of the beam having its incoherent tune shift $\delta\nu$ and actions $J_x$, $J_y$, $J_z$. Linearized Vlasov equation for this beamlet can be presented similar to Eq. (4):

$$\nu\mathbf{x} = \delta\nu\mathbf{x} + \hat{\mathbf{S}}\cdot\mathbf{x} + \hat{\mathbf{M}}\cdot\mathbf{X}, \tag{18}$$

where $\nu$ stays for a perturbed mode frequency of the nonlinear system, matrix $\hat{\mathbf{M}}$ combines all the collective response matrices generating a coherent field $\hat{\mathbf{M}}\cdot\mathbf{X}$, and the coherent vector $\mathbf{X}$ is a beam-average of the beamlet perturbations $\mathbf{x}$.

From here, the beamlet perturbation $\mathbf{x}$ can be expressed in terms of the collective perturbation $\mathbf{X}$. After the beam-averaging, $\langle\ldots\rangle$, this yields

$$\mathbf{X} = \left\langle\left[(\nu - \delta\nu + io)\hat{\mathbf{I}} - \hat{\mathbf{S}}\right]^{-1}\right\rangle\cdot\hat{\mathbf{M}}\cdot\mathbf{X}, \tag{19}$$

where $o$ stands for a positive infinitesimally small number showing how to bypass the pole. Assuming that a solution of this set of equations is close to the unperturbed one, with $\delta\nu = 0$, the theorem of Eq. (17) can be applied, leading to a dispersion equation for the perturbed mode tune $\nu_k$:

$$\mathbf{Y}_k^{\dagger}\cdot\left\langle\left[(\nu_k - \delta\nu + io)\hat{\mathbf{I}} - \hat{\mathbf{S}}\right]^{-1}\right\rangle\cdot(q_k\hat{\mathbf{I}} - \hat{\mathbf{S}})\cdot\mathbf{X}_k = 1, \tag{20}$$

where $\mathbf{X}_k$ and $\mathbf{Y}_k$ are the unperturbed eigenvectors,

$$\begin{aligned}(\hat{\mathbf{M}} + \hat{\mathbf{S}} - q_k\hat{\mathbf{I}})\cdot\mathbf{X}_k &= 0;\\ (\hat{\mathbf{M}}^{\dagger} + \hat{\mathbf{S}} - q_k^{*}\hat{\mathbf{I}})\cdot\mathbf{Y}_k &= 0;\end{aligned} \tag{21}$$

the normalization condition $\mathbf{Y}_k^{\dagger}\cdot\mathbf{X}_k = 1$ was assumed. Since all the matrices of Eq. (20) are diagonal, the dispersion equation (20) can be further simplified. For every unperturbed eigenvalue *q*, the corresponding mode tune with Landau damping taking into account can be found from the following dispersion equation

$$\begin{aligned}&-\sum_l (q-l)\int\frac{Y_l^{*}(J_z)X_l(J_z)J_x\dfrac{\partial F}{\partial J_x}}{\nu - l - l\delta\nu_s - \delta\nu_x + io}d\Gamma = 1;\\ &\mathbf{Y}^{\dagger}\cdot\mathbf{X} = \int Y_l^{*}(J_z)X_l(J_z)F_z(J_z)dJ_z = 1;\\ &d\Gamma \equiv dJ_x dJ_y dJ_z; \quad \int F d\Gamma = \int F_z(J_z)dJ_z = 1.\end{aligned} \tag{22}$$

Here, $F = F(J_x, J_y, J_z)$ and $F_z(J_z)$ are total and longitudinal normalized phase space densities as functions of corresponding actions, while $\delta\nu_s = \delta\nu_s(J_z)$ and $\delta\nu_x = \delta\nu_x(J_x, J_y, J_z)$ are incoherent anharmonicities of the normalized synchrotron and betatron tunes. Note that the beam averaging $\langle\ldots\rangle$ is expressed as

$$\langle\ldots\rangle = -\int\ldots J_z\frac{\partial F}{\partial J_z}d\Gamma, \tag{23}$$

not by a naively expected $\int\ldots F d\Gamma$. For an ensemble of non-harmonic oscillators, that rule was derived by H.G. Hereward [13], and explained by him as resulting from an incoherent frequency shift when the coherent field acts on a given particle. In general, Eq. (22) tells that the mode tune shift $\nu$ is determined not only by its harmonic approximation *q*, the incoherent spectrum $\delta\nu$, and the phase space density F, but also it depends on the eigenvectors $\mathbf{X}$ and $\mathbf{Y}$. The latter is expectable: a contribution to Landau damping of particles with the longitudinal action $J_z$ has to be proportional to the wave amplitude at this action. Note that Eq. (22) assumes the sought-for mode tune $\nu$ be located at the upper complex half-plane, while at the lower half-plane the left-hand side (LHS) of Eq. (22) has to be understood as an analytical continuation. In many practical cases, it is sufficient to tell if the system is stable or not under given conditions, while a specific value of the mode tune is not so important. For that purpose, it is reasonable to consider the LHS of Eq. (22) as a map of the real axis $\nu$. When $\nu$ is running from $-\infty$ to $\infty$, the LHS follows a closed curve in its complex plane, sometimes referred as a lotus. The system is stable if, and only if, the lotus does not cover the point $1 + 0i$.

In case of the weak head-tail, when the eigenvalue *q* is close to its nearest integer $l_q = \mathrm{Round}(q)$, $|q - l_q| \ll 1$, and the synchrotron tune spread can be neglected, the longitudinal integration can be performed, resulting in the well-known dispersion equation [6]:

$$\begin{aligned}&-\Delta q\int\frac{J_x\partial F/\partial J_x}{\Delta\nu - \delta\nu_x(J_x, J_y) + io}dJ_x dJ_y = 1;\\ &\Delta q \equiv q - l_q; \quad \Delta\nu \equiv \nu - l_q.\end{aligned} \tag{24}$$

For this, and apparently only for this case, the dispersion relation, and thus a stability condition, does not depend on the eigenvectors. Then, the stability condition can be expressed in terms of the stability diagram independent of the eigensystem [14, 15]. Namely, Eq. (24) can be presented as

$$\begin{aligned}&\Delta q = D(\Delta\nu);\\ &D(\Delta\nu) \equiv -\left(\int\frac{J_x\partial F/\partial J_x}{\Delta\nu - \delta\nu_x + io}d\Gamma\right)^{-1}.\end{aligned} \tag{25}$$

An expression $D(\Delta\nu)$ can be considered as a map of the real axis in the complex plane $\Delta\nu$ onto a complex plane *D*. As such, it is conventionally referred as the stability diagram. The diagram curve goes along the real axis of the *D*-plane, $D(\Delta\nu)\approx\Delta\nu$, when $|\Delta\nu|\gg\sqrt{\langle\delta\nu^2\rangle}$. The width and the height of the stability diagram are determined by the averaged nonlinearities $\delta\nu$ for the phase space density *F*. The same stability diagram is valid for all the modes: as soon as the eigenvalue $\Delta q$ is located above the diagram, the mode is unstable; otherwise, it is suppressed by Landau damping.

For Gaussian transverse distribution, and with negligible spread of the synchrotron frequencies, the 2D dispersion integral was found by R. Gluckstern [6]:

$$\int_0^\infty\!\!\int_0^\infty \frac{x\exp(-x-y)dxdy}{\nu-ax-by+io}= -\frac{a-b+(b-\nu(1-b/a))\exp(-\nu/a)\,\mathrm{Ei}(\nu/a)}{(a-b)^2}+ \frac{b\exp(-\nu/b)\,\mathrm{Ei}(\nu/b)}{(a-b)^2}- \pi i\frac{\left|-(b-\nu(1-b/a))\exp(-\nu/a)\theta(\nu/a)+b\exp(-\nu/b)\theta(\nu/b)\right|}{(a-b)^2};$$

$$\mathrm{Ei}(z)\equiv -P.V.\int_{-z}^{\infty}(e^{-t}/t)dt\,. \qquad (26)$$

Here *P.V.* stays for the principle value and $\theta(z)$ is the Heaviside theta-function. Stability diagrams for distribution functions $F(J_x,J_y)\propto(1-(J_x+J_y)/a)^n$ are discussed in Ref. [16].

For the LHC impedance model, the highest coherent tune shifts are comparable to the synchrotron tune if the damper is turned off [17], making the weak head-tail approach to Landau damping (24), (25) marginally applicable in this area of the parameters. However, when the head-tail phase and the gain are sufficiently large, the most critical coherent tune shifts *q* are reduced several times [5], allowing to rely on the weak-head tail stability diagram (25) without any visible loss of accuracy.

## BENCHMARKING

Above, the main ideas and formulas of the NHT code are described. Some NHT results for LHC were compared with the Air-Bag Averaged (ABA) Vlasov solver [18] and BeamBeam3D tracking code [4, 19], showing a good agreement for all the examined cases.

In Ref. [19], instability growth rate computed from the BeamBeam3D tracking simulations for LHC parameters is presented as a function of the chromaticity and damper gain. For those simulations, a single 3D-Gaussian bunch per beam and single IP were assumed. The intensity and collision parameters were taken close to the end of the beta-squeeze case. Namely, 10 rms beam radius of the beam-beam separation was assumed, and the computed beam-beam long-range kick was additionally enhanced by a factor of 10, thus simulating 10 identical long-range collisions instead of one. The IP optics was taken as perfectly round, all the octupoles were zeroed, the potential well was supposed to be ideally parabolic, and the doubled nominal impedance of the LHC was implemented.

To make a comparison, NHT computations were fulfilled for the same conditions. Without octupoles and longitudinal nonlinearity, the only source of the Landau damping is a long-range beam-beam octupole term yielding the incoherent tune shift per collision, which can be presented as

$$\delta\nu_x=\frac{3\xi}{2\tilde\rho^2}\frac{J_x-2J_y}{\varepsilon_n};\qquad (27)$$
$$\langle J_x\rangle=\langle J_y\rangle=\varepsilon_n\,,$$

where the beam-beam parameters $\xi,\tilde\rho$ are the quadrupolar incoherent beam-beam tune shift and normalized separation, introduced at Eq.(11).

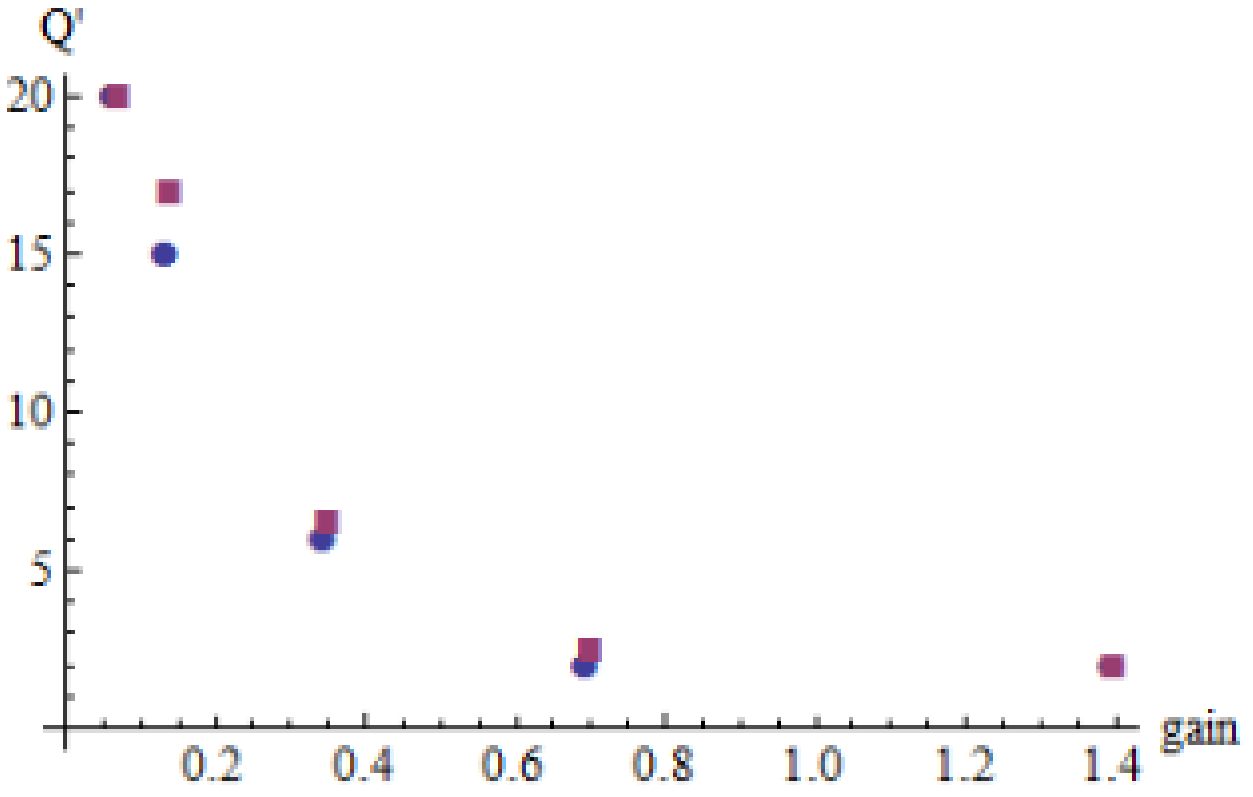

Figure 2: Threshold chromaticity $Q'$ versus the damper gain for BeamBeam3D tracking (circles) [17] and NHT solutions (squares).

In Ref. [19], the instability threshold is shown as a threshold chromaticity $Q'$ for certain values of the damper gain. In Fig. 2, these results of BeamBeam3D are presented together with corresponding NHT ones.

To appreciate the agreement between the two sets of results, one has to take into account that at high chromaticity, $Q'\gtrsim 10$, the stability condition is barely sensitive to $Q'$ (see the following section), which significantly amplifies initially small computational errors when the chromaticity is that high. Note that the NHT data at Fig. 2 reflect eigenvalue computation with beam-beam on with the following analysis of the stability diagram. Figure 3 shows BeamBeam3D single bunch, no damper, no octupoles growth rates [19] compared with corresponding NHT eigenvalues $\mathrm{Im}\, q$ for the most unstable mode, demonstrating agreement within few percent or better.

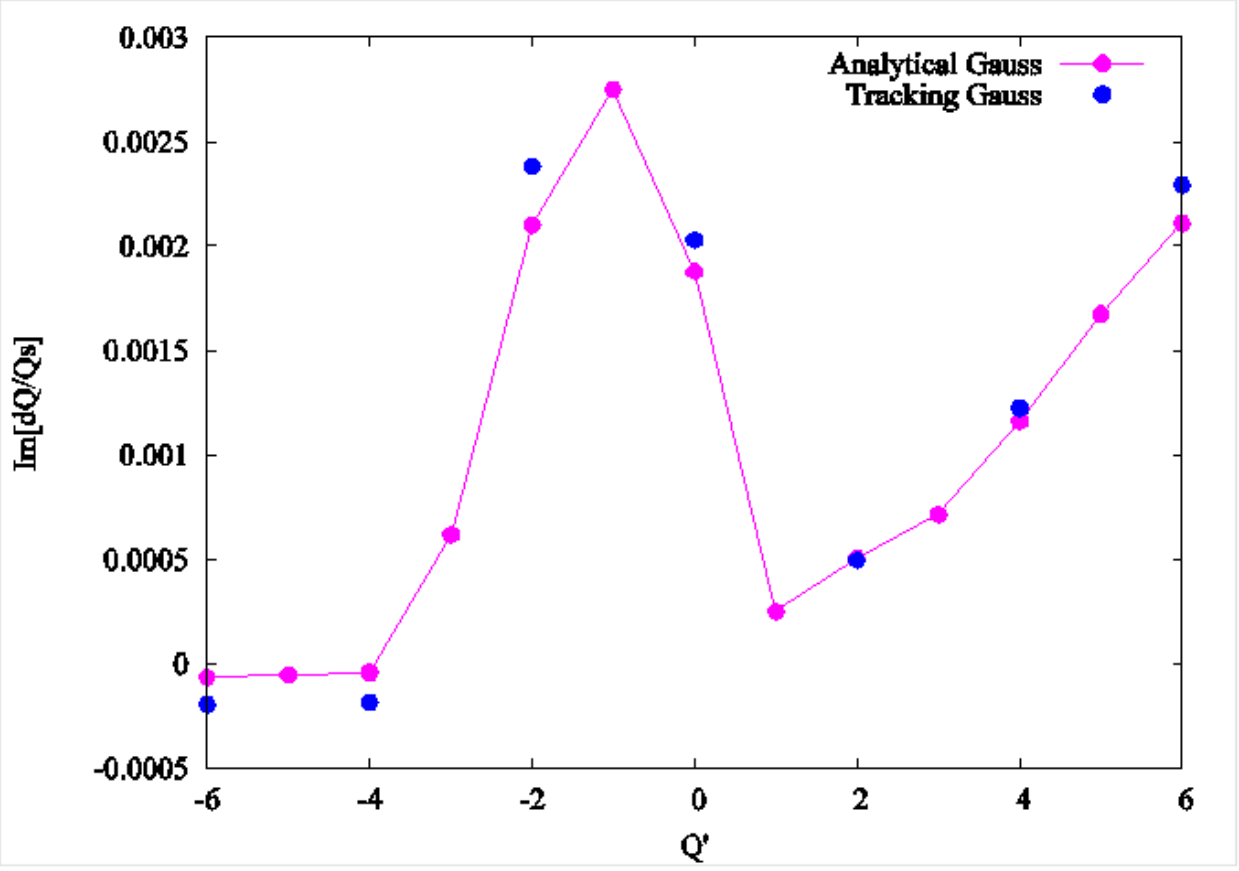

Figure 3: BeamBeam3D (blue) and NHT (pink) growth rates versus chromaticity for a single bunch, no octupoles and no damper.

## LHC: EIGENSISTEM

In the rest of this paper, various capabilities of the Nested Head-Tail program are demonstrated for the LHC at 4 TeV. All the computations are done for the horizontal degree of freedom. LHC horizontal wake and impedance functions are presented in Fig. 4 and 5, as they were provided to the author by N. Mounet [18, 20]. According to this model, vertical wake and impedance are similar to horizontal, being slightly smaller. Hereafter, this impedance model is referred to as a nominal one. Various beam-based measurements at the LHC are showing, though, that actual impedances are ~2-3 times higher than the nominal ones (see Ref. [17] and following sections). So far, a reason for this discrepancy is unknown. Strictly speaking, it has to be supposed that not only a scale of the impedance is higher, but its frequency dependence may be considerably different from the model as well. However, due to a lack of knowledge about the real impedance of the LHC, the NHT computations are normally performed just with the doubled nominal impedance and wake functions. Figures below assume $N_b = 1.5 \cdot 10^{11}$ protons per bunch, $50$ ns of the bunch separation, the normalized rms emittance $\varepsilon_n = 2\ \mu\mathrm{m}$, the synchrotron tune $Q_s = 2.3 \cdot 10^{-3}$, and the rms bunch length $c\bar{\tau} = 9.4\ \mathrm{cm}$. In this section, properties of the eigensystem are discussed; the next one is devoted to the stability analysis.

### *Single Beam*

Figures 6 and 7 show an example of horizontal eigenvalues of Eq. (4) for the specified beam at the chromaticity $Q_x' = 15$ , yielding the rms head-tail phase $\bar{\chi} \equiv Q_x' \omega_0 \bar{\tau} / \eta = 1.0$; the damper was assumed to be flat bunch-by-bunch. The gain value $g = 1.4$ is a standard Run I setting corresponding to 50 turns of the damping time for the 50 ns beam.

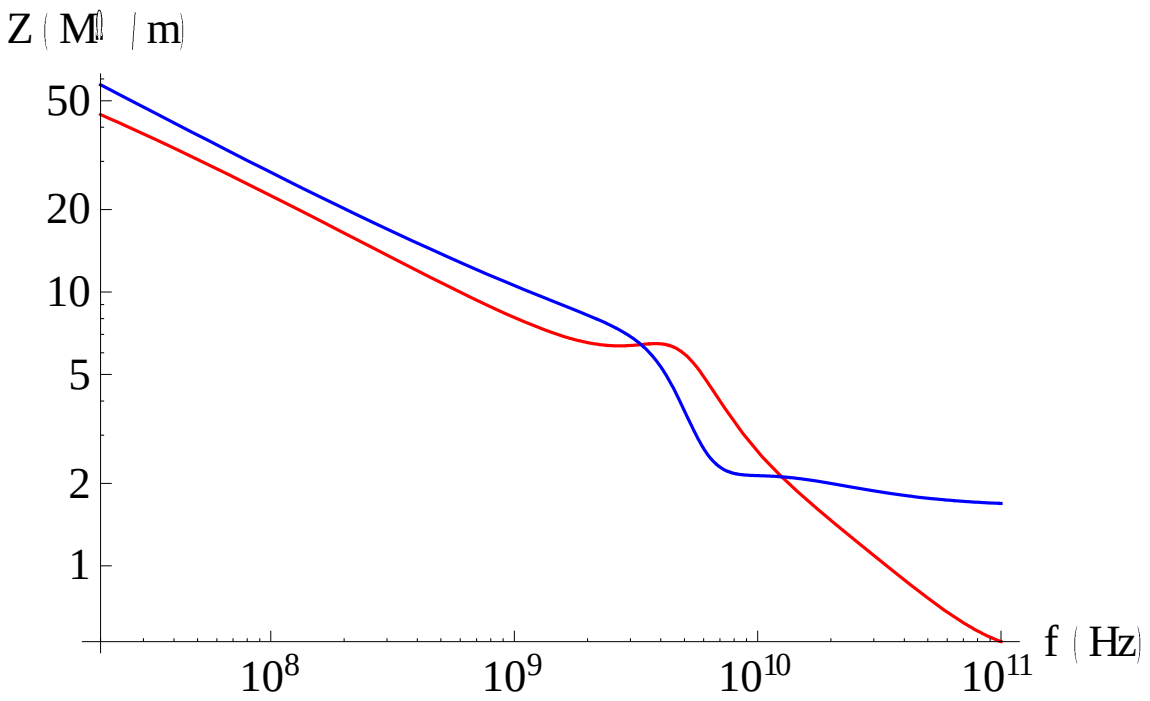

Fig. 4: Horizontal impedance, nominal [20].

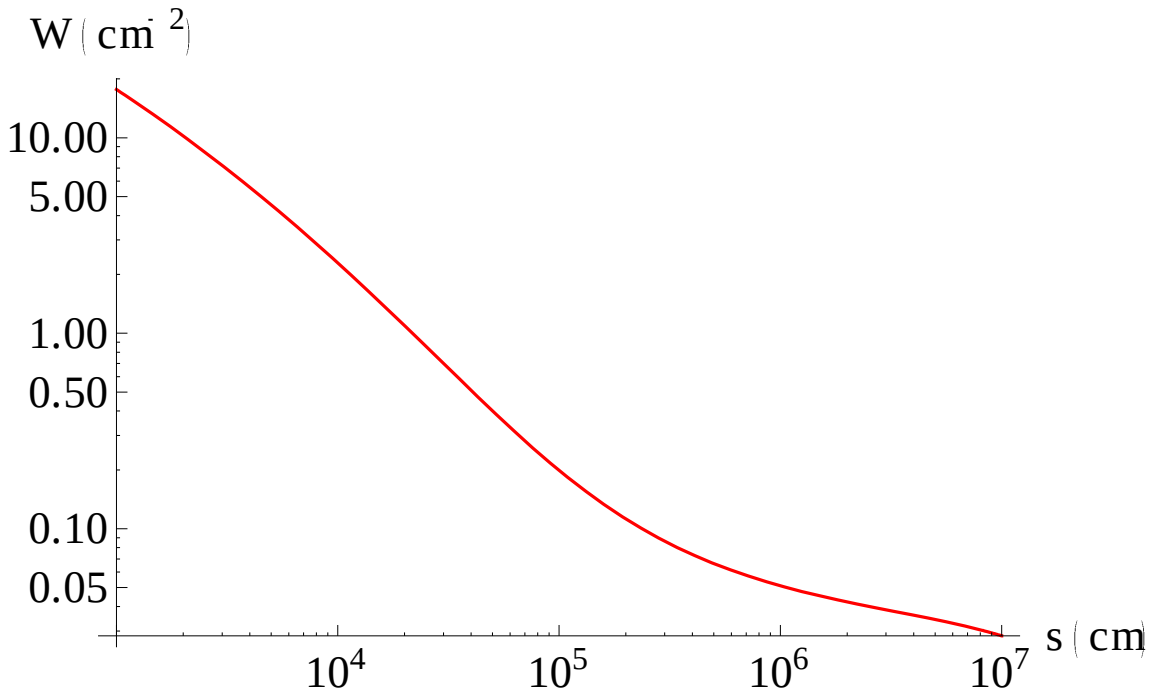

Fig. 5: Horizontal wake function, nominal [20].

Figure 8 offers a 3D plot for maximal growth rate $\mathrm{Im}\, q_* \equiv \mathrm{Max}[\mathrm{Im}(q)]$ versus the gain and chromaticity. A significant difference between positive and negative chromaticity is not surprising, but non-monotonic dependence over the gain at the negative chromaticity area was not expected. Note a stable area at moderate gain and slightly negative chromaticity.

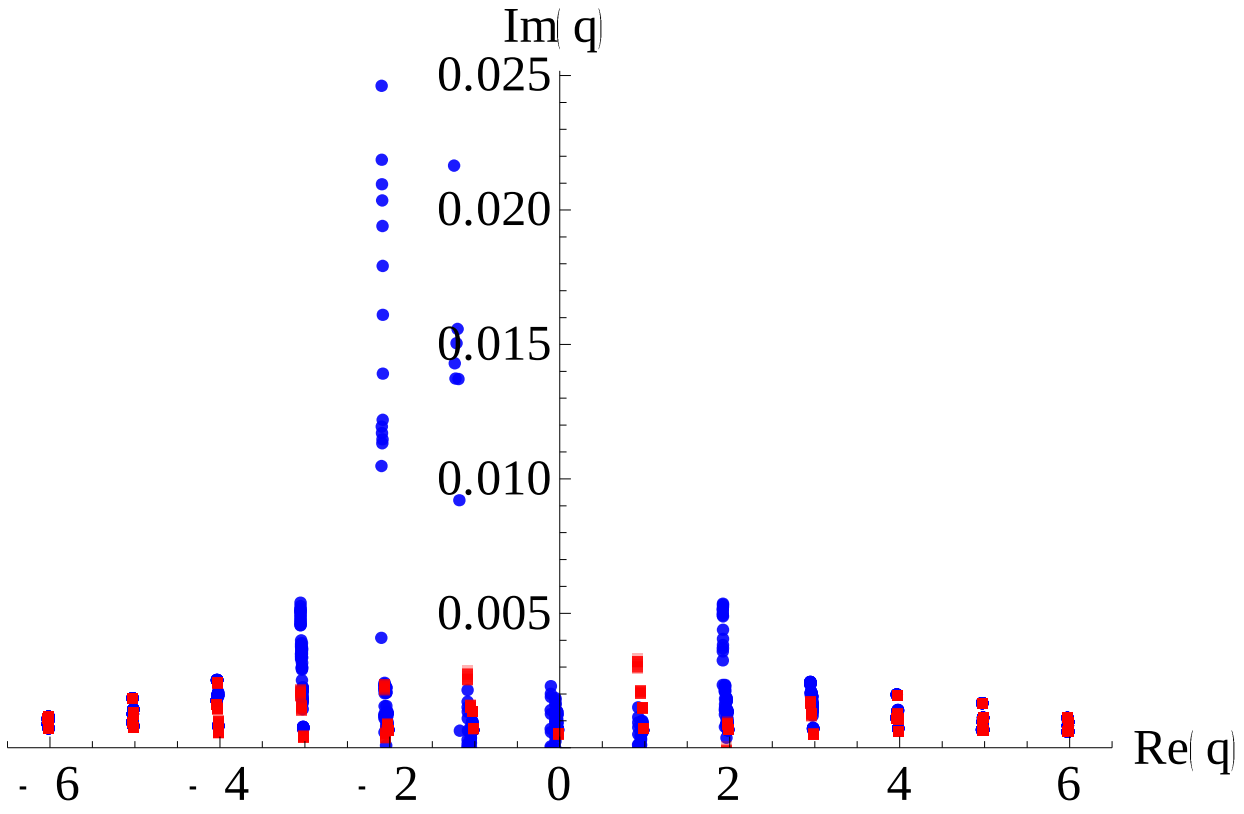

Figure 6: Horizontal eigenvalues of a standard single beam with the damper gain $g = 1.4$ and chromaticity

$Q'_x = 15$ (red). The blue dots show the same case with the damper off.

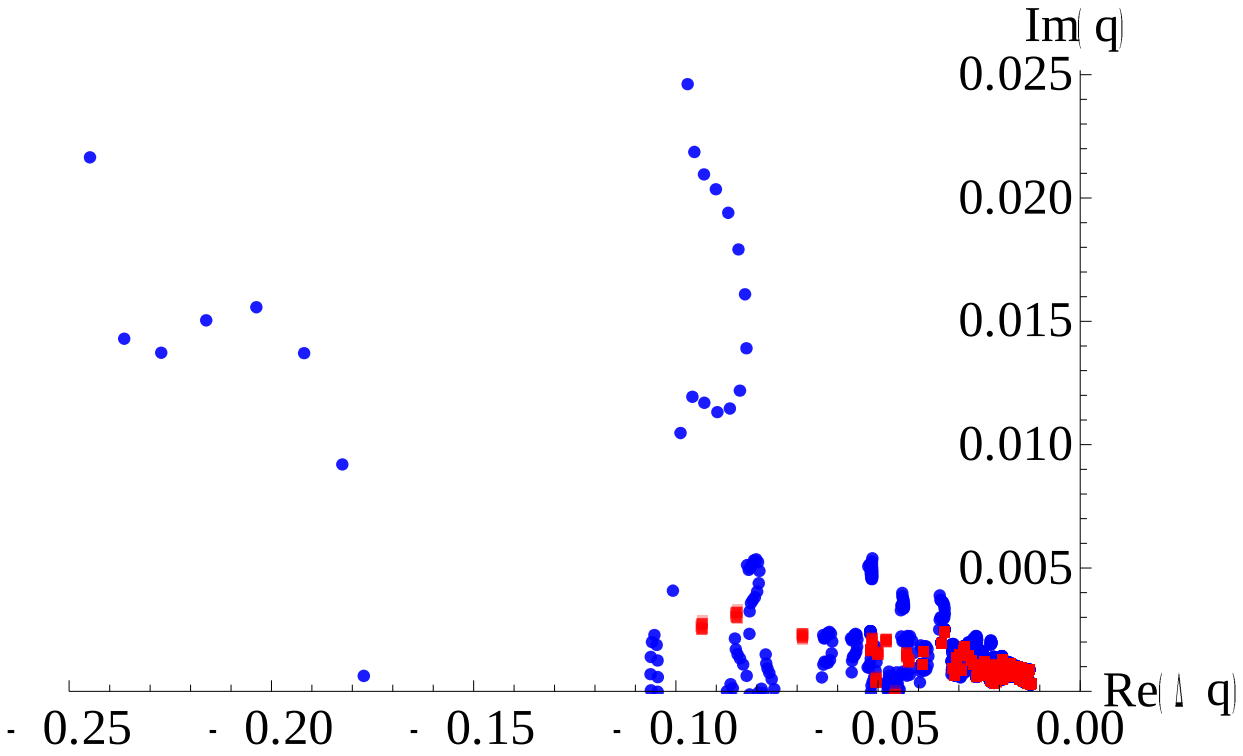

Figure 7: The same as the figure above, but for the fractional eigenvalues $\Delta q = q - \mathrm{Round}[q]$ entering into the stability condition, Eq. (24).

Figure 9 presents intensity-chromaticity scan of the maximal growth rate, assuming the standard gain $g = 1.4$. The coefficient $K_Z$ is intensity scaling parameter, defined as impedance time bunch population versus the nominal value of this product. All the plots above correspond to $K_Z$ =2.

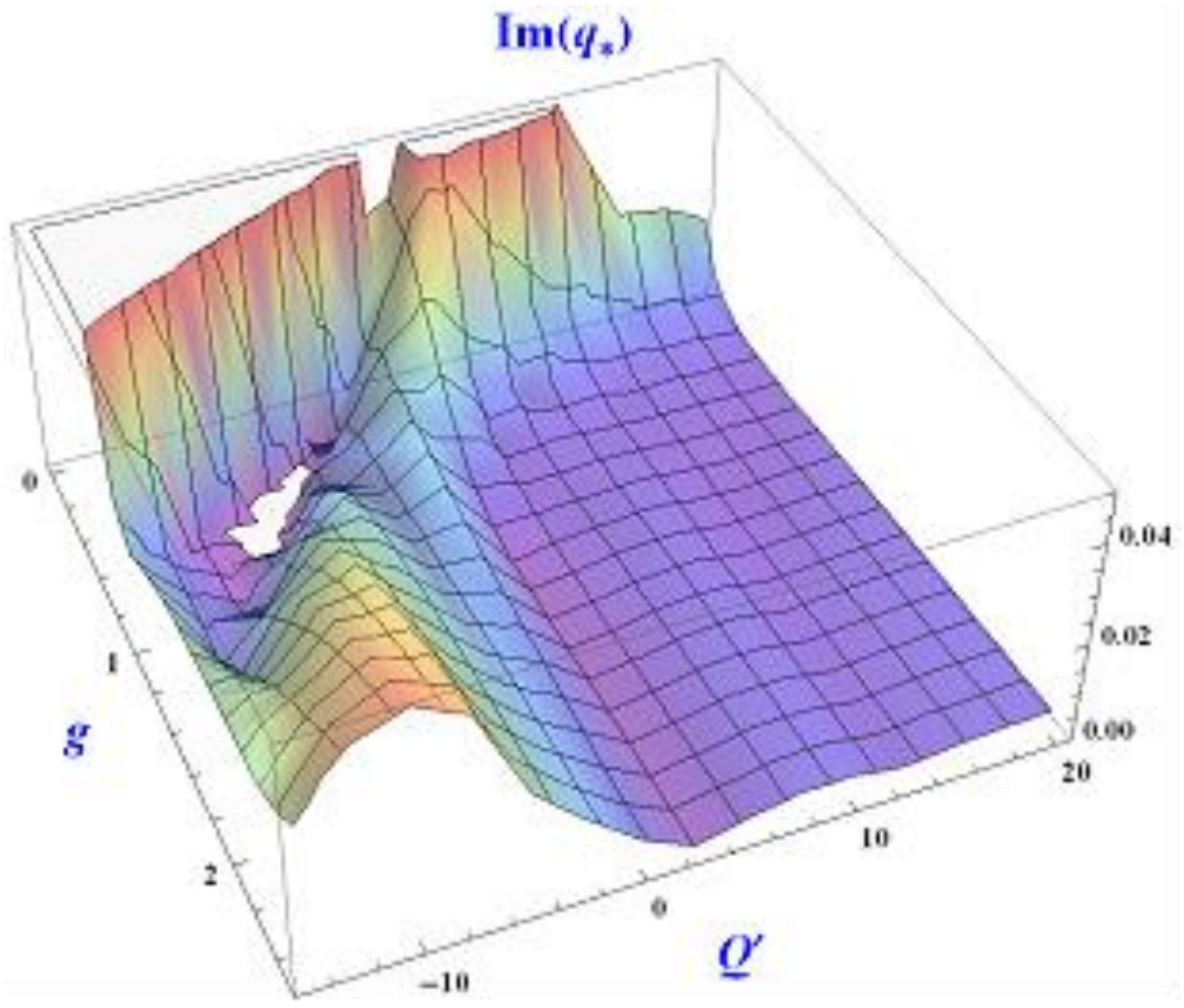

Figure 8: Highest growth rate versus gain and chromaticity

## *Two Beams*

Similar results for two beams in the LHC, Eq. (14), are presented below with long-range beam-beam parameter, Eq.(11), $\xi = 2.5 \cdot 10^{-3}$ per interaction region (IR), as it is at the end of the LHC beta-squeeze, and the inter-IR beam1-beam2 phase difference $\psi = 90^\circ$. While no-damper growth rates (blue) are about doubled by the beam-beam interaction, they are changed much less for the standard gain $g = 1.4$. This property of the coherent spectrum is a consequence of the "damper theorem", see Ref. [8].

Up to this point, all the plots show eigenvalues for pure linear particle motion. The next section takes into account octupole transverse nonlinearity causing Landau damping and responsible for certain instability thresholds. Note that maximal growth rates $q_*$ shown in several plots above are not generally proportional to the octupole strength, required for stabilization, since this strength depends not only on the growth rate, but on the tune shift $\mathrm{Re}(\Delta q)$ as well.

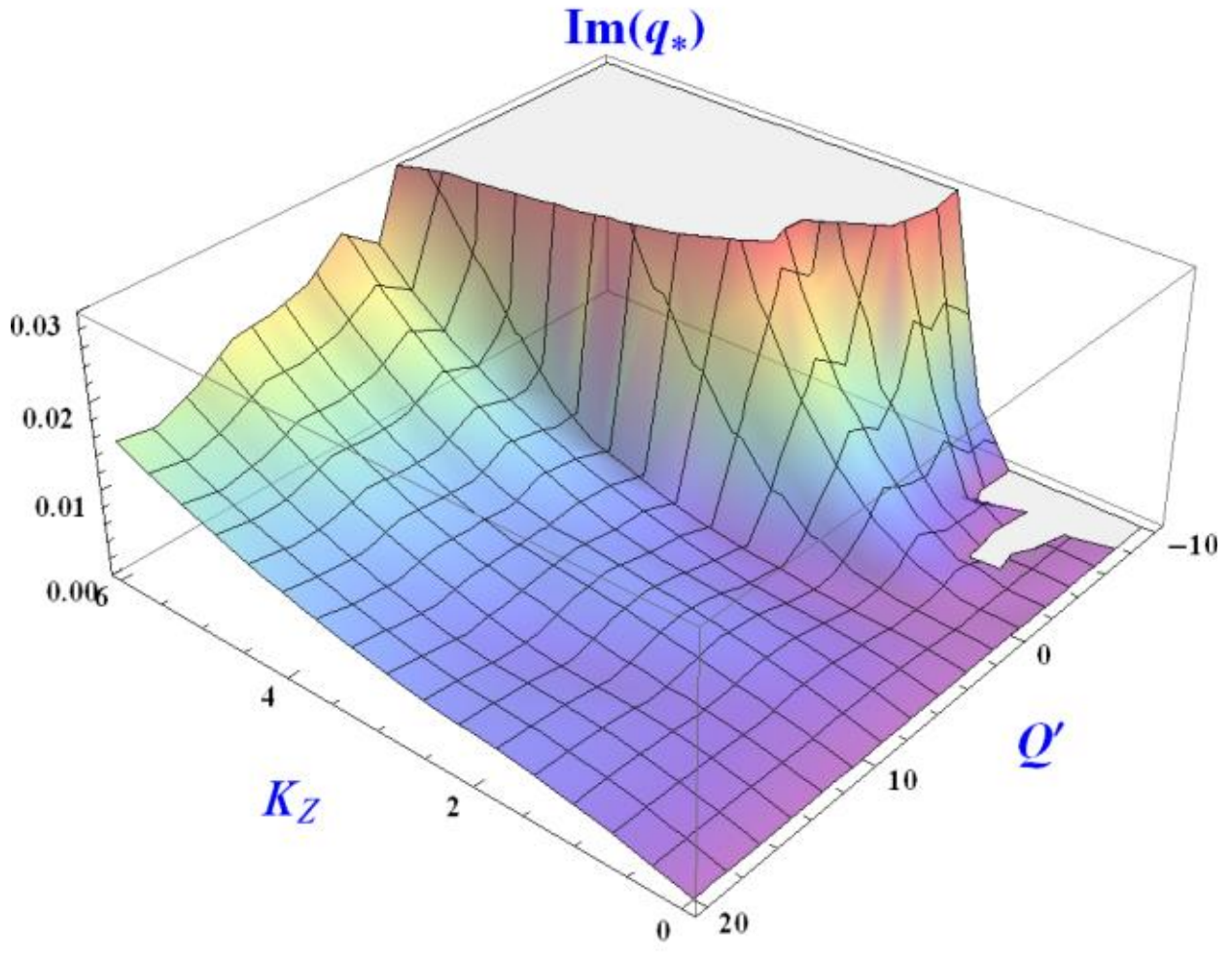

Figure 9: Maximal growth rate versus intensity and chromaticity at the gain $g = 1.4$ .

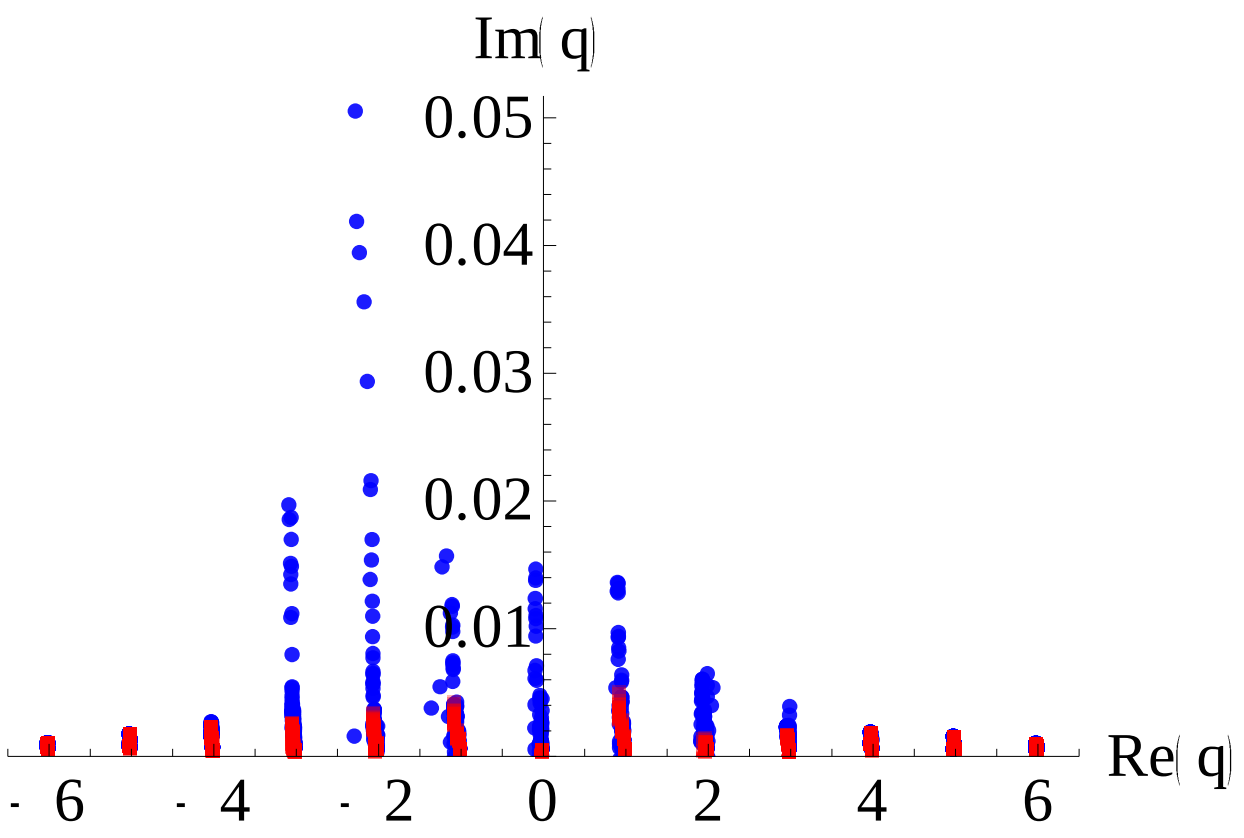

Figure 10: Two-beam eigenvalues for the same single-beam parameters and colour convention as Fig. 6.

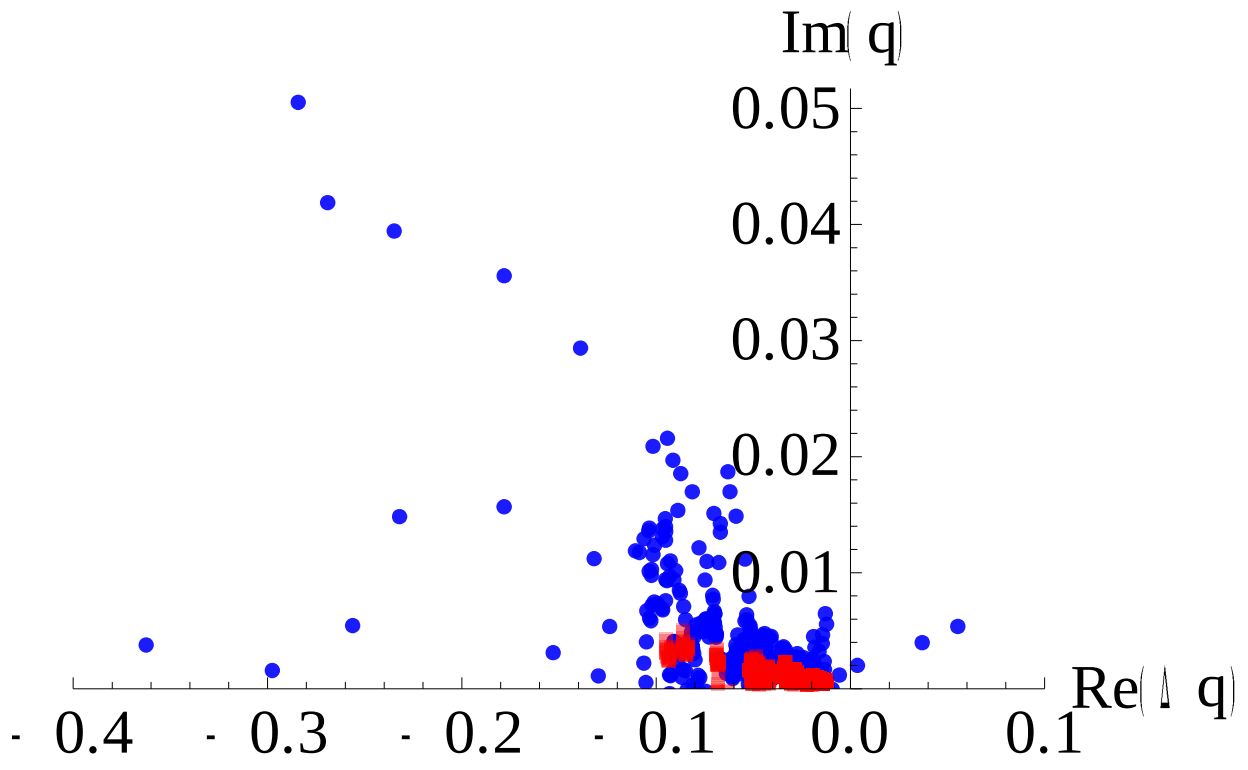

Figure 11: Same as the previous figure, but for the fractional eigenvalues.

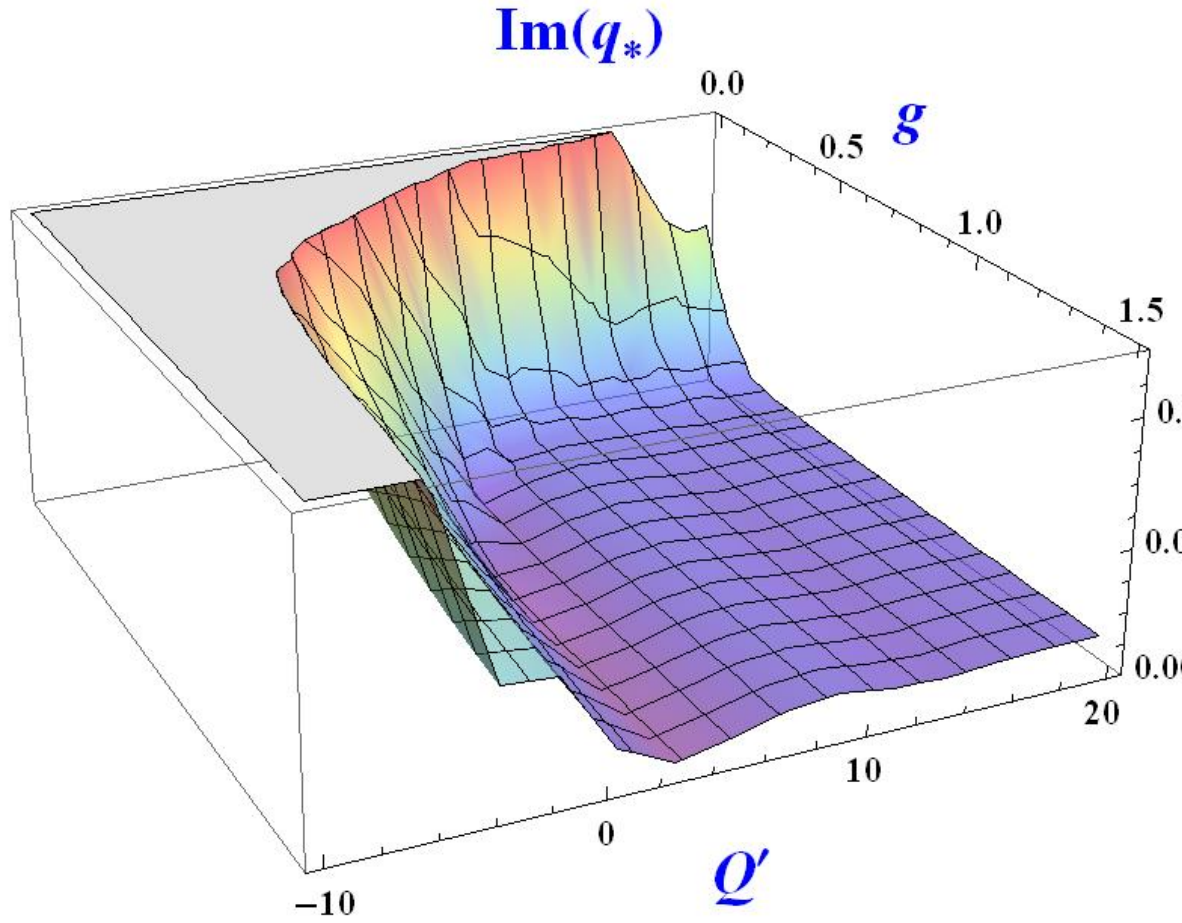

Figure 12: Maximal growth rate for two beams versus gain and chromaticity.

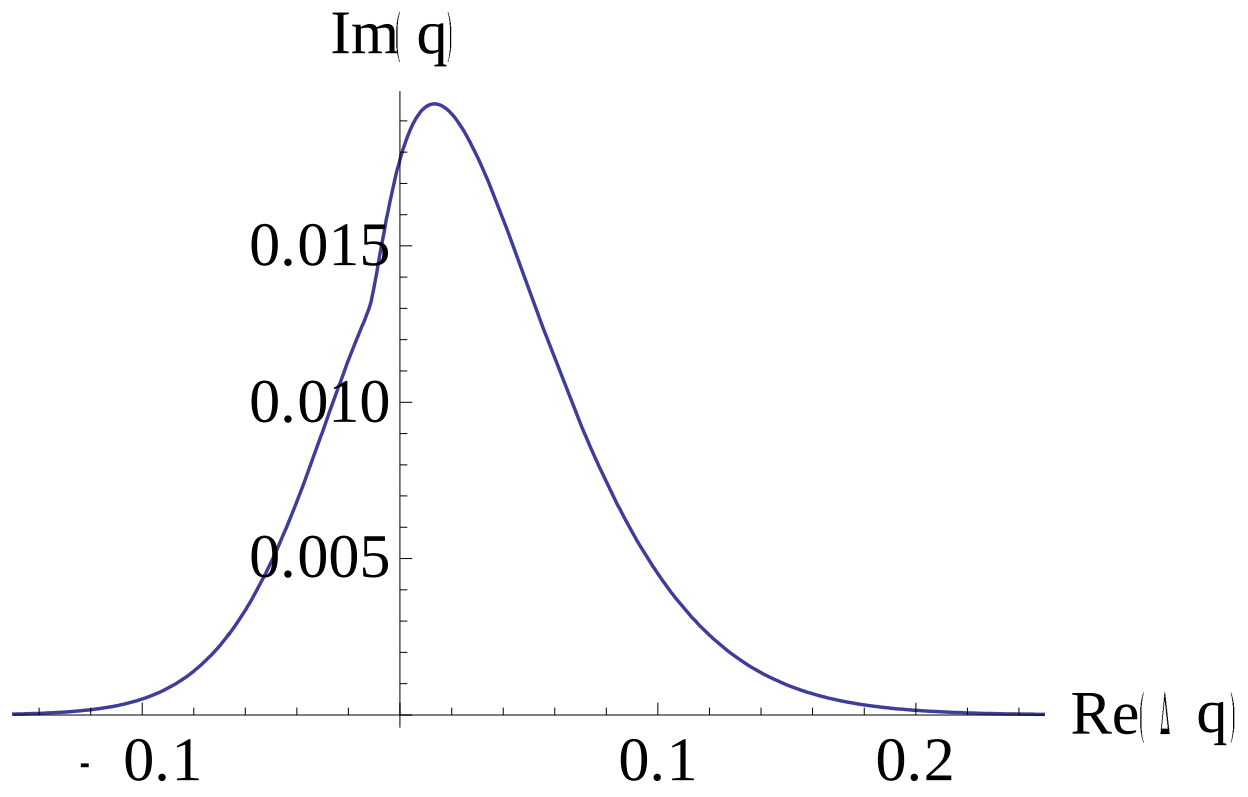

Figure 13: Stability diagram for LHC octupoles at their current $I_o = +100\text{A}$.

## LHC: INSTABILITY THRESHOLDS

For Landau damping, two families of octupoles are installed in the LHC, normally fed with the same currents $I_o$. According to Ref. [20, 21], the incoherent tune shifts $\boldsymbol{\delta\nu} \equiv \left(\delta\nu_x, \delta\nu_y\right)^{\mathrm{T}}$ introduced by the LHC octupoles fed with current $I_o = +100\text{A}$, with both emittances $\varepsilon = 2\mu\text{m}$ are as follows:

$$\begin{aligned} &\boldsymbol{\delta\nu} = \hat{\mathbf{A}} \cdot \mathbf{J} / \varepsilon\,; \quad \mathbf{J} \equiv (J_x, J_y)^{\mathrm{T}}; \\ &\hat{\mathbf{A}} = \begin{pmatrix} a_{xx} & a_{xy} \\ a_{yx} & a_{yy} \end{pmatrix}; \\ &a_{xx} = a_{yy} = 1.8 \cdot 10^{-2}; \\ &a_{xy} = a_{yx} = -1.3 \cdot 10^{-2}; \end{aligned} \tag{28}$$

Substitution of that in the dispersion relation, Eq. (25), for a Gaussian beam yields the stability diagram of Fig. (13). For every given chromaticity and gain in a grid, the threshold octupole current was found, i.e. such a current when all the fractional eigenvalues lie below the stability diagram, except one sitting on this curve. The threshold octupole currents $I_o$ as functions of chromaticity and gain are presented in Fig. 14 (single beam, single bunch), Fig. 15 (one 50ns beam) and Fig. 16 (two 50ns beams). Computational time for each one of these plots is about a half an hour at PC laptop with Intel(R) Core(TM)2 Duo CPU, 2.5GHz processor. While these three plots are very different at not a very high gain and negative chromaticity, they are close at high gain, demonstrating validity of the "damper theorem", Ref. [8]. A special practical interest suggests a plateau $I_o \cong 120-140\text{A}$ at positive chromaticity region, where stability is provided by a relatively low octupole current insensitive to errors in the chromaticity.

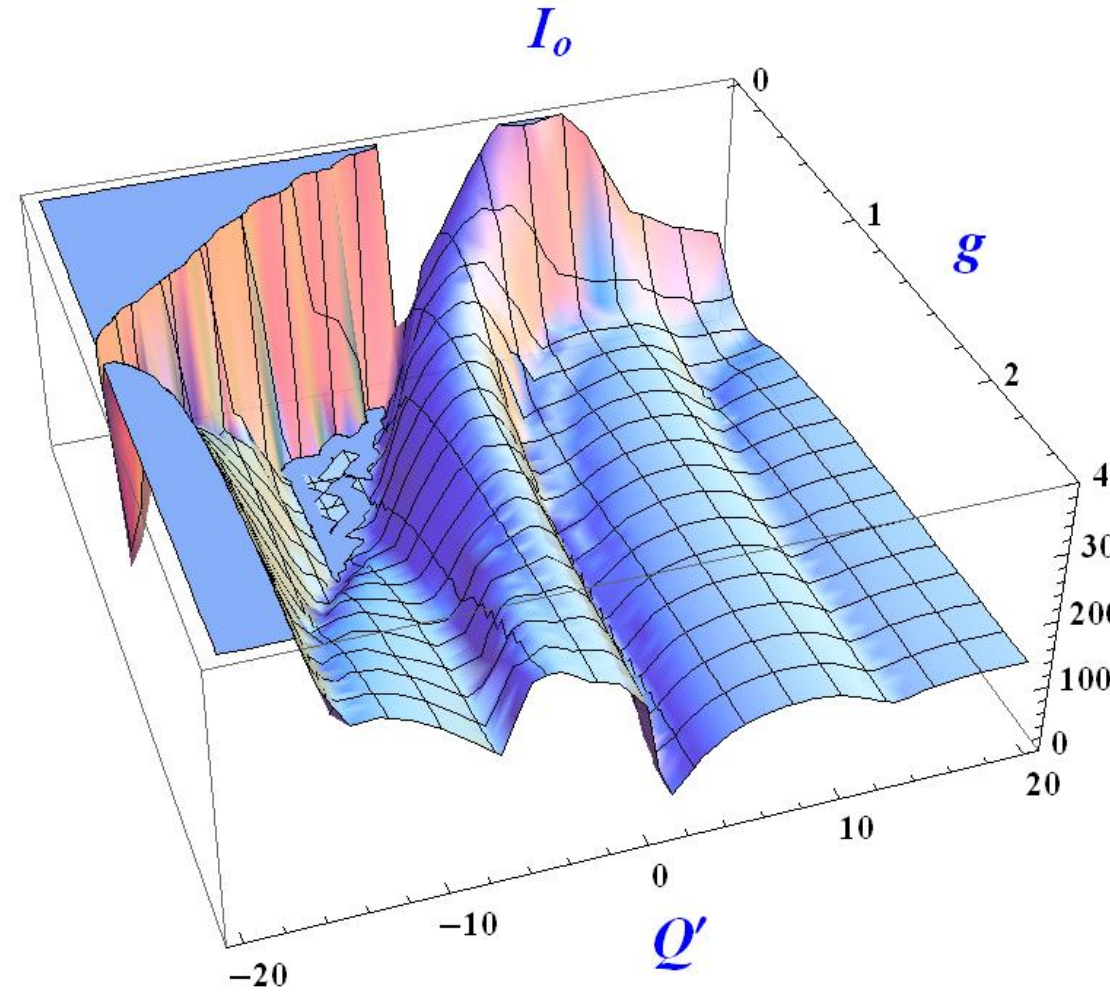

Figure 14: Threshold octupole current [A] at positive polarity, for a single nominal bunch in the LHC and doubled nominal impedance.

Note that Figs. (14-16) do not take into account beam-beam nonlinearity. At first glance, Eq. (27) could be used for the octupole component of the long-range tune shift to be taken into account together with the LHC Landau octupoles. To do this, one has to double this tune shift, since there are two interaction regions, and divide it back by a factor of two for the first or the last bunch of the batch (pacman bunches), so that for the pacman LHC bunches Eq. (27) gives the result. However, this result would not be normally correct, since in the reality the beams are initially separated in the orthogonal plane as well, which is not taken into account by Eq. (27). Analysis of the stability diagram of Gaussian beams provided by Ref. [22] demonstrates that with this additional beam-beam separation, the beam-beam nonlinearity is approximately equivalent to 100A of the Landau octupoles at the end of the beta-squeeze. For positive polarity of the octupoles, these 100A of beam-beam nonlinearity go with the same sign as the octupoles, thus increasing Landau damping. For negative polarity, the two nonlinear contributions are of the opposite sign, leading to a collapse of the stability diagram at certain beam-beam separation. When this was realized [23], the initially negative octupole polarity has been inverted.

Single-beam measurements of the instability thresholds made at the high gain and high chromaticity plateau never exceeded 200A of the Landau octupoles [17]. Compared with Fig. 15, it leads to the conclusion that the effective single-bunch impedance of the LHC should be 2.5-3 times higher than the nominal one of Fig. 4. Figure 16 shows the effective octupole current required for stabilization of two LHC beams seeing each other at the end of the beta-squeeze. The effective octupole current is the sum of the Landau octupole current and a contribution of the beam-beam nonlinearity expressed in terms of the equivalent octupole current. According to Ref. [22], the oncoming beam contributes 100A to the effective octupole current for the pacman bunches and twice more for the central ones at the end of the squeeze. It follows then, that about 100A of the Landau octupoles should be sufficient for the stabilization, assuming the machine is operated at the high gain and high chromaticity plateau of Fig. 16, as it normally was. Contrary to this conclusion, at the end of the squeeze a transverse instability was permanently observed, notwithstanding the octupole current was kept at its maximum of 550A [24].

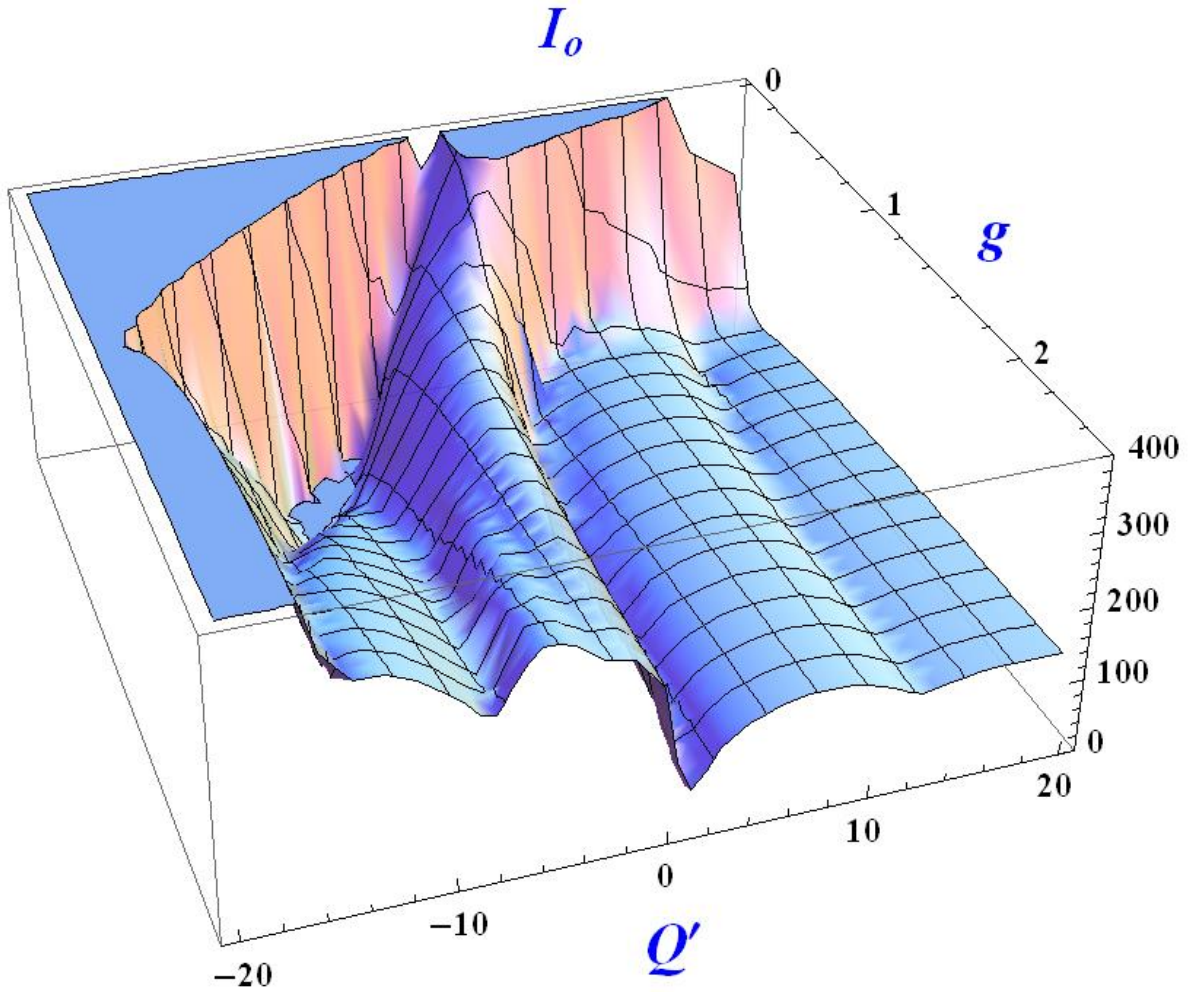

Fig. 15: Same as above, but for a full 50ns single beam.

An initial idea that this instability is driven by the coherent (strong-strong) beam-beam effect or some hidden two-beam impedance was refuted by these NHT computations [25], confirmed later by a dedicated LHC beam experiment [26]. To explain this instability, a hypothesis of three-beam instability, or beam-beam-beam effect was suggested, where the third beam is an electron cloud accumulated in a high-beta area of the main interaction regions [27].

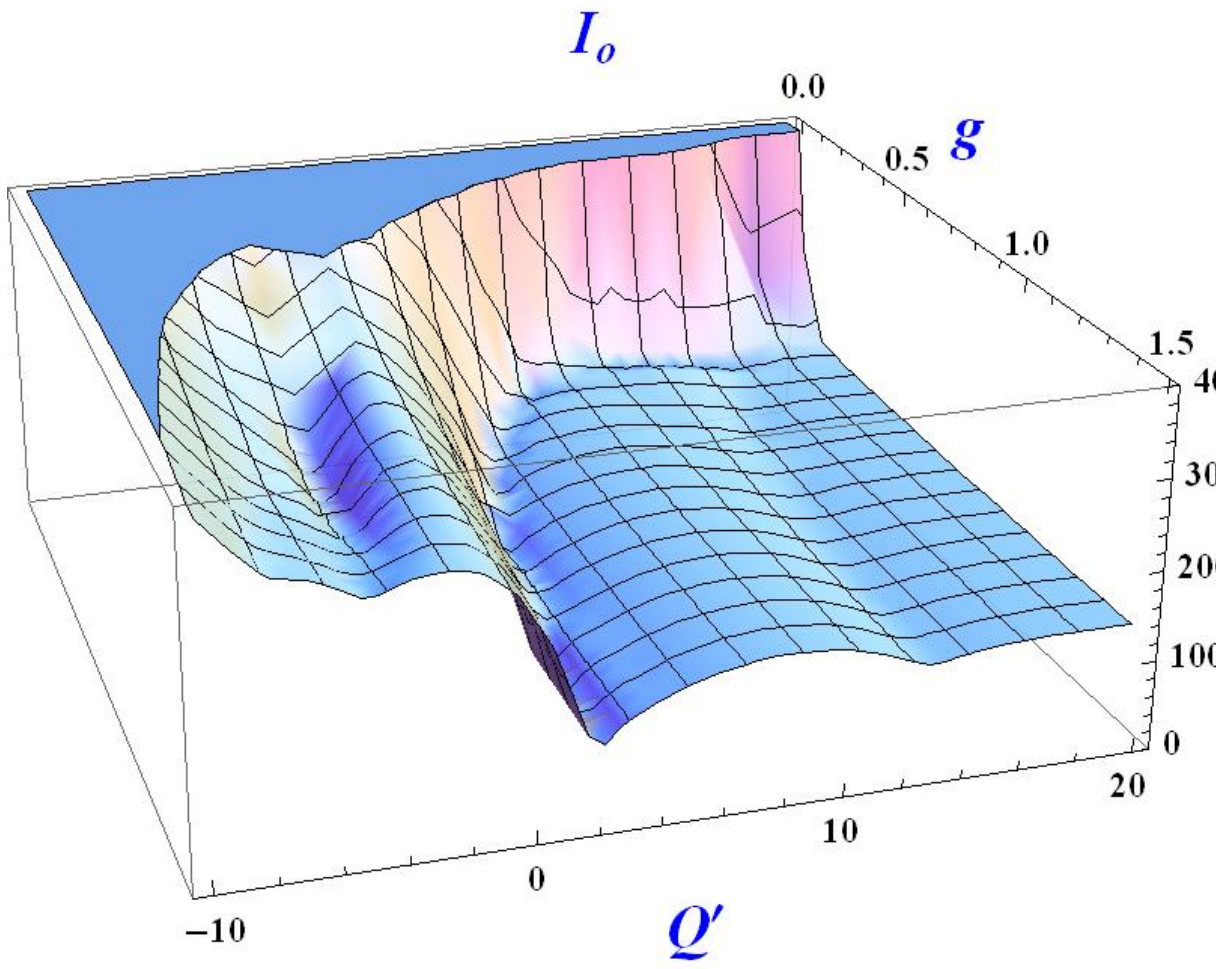

Fig. 16: Same as above, but for two 50ns single beams at the end of the beta-squeeze separation.

## SUMMARY

This paper describes Nested Head-Tail Vlasov solver effectively used for high energy beams in LHC, where radial modes, couple-bunch modes, feedbacks, beam-beam effects and nonlinearities responsible for Landau damping are accurately handled. Main advantage of that solver against macroparticle tracking codes is many orders of magnitude shorter CPU time, which allows a fast and efficient analysis of that complicated system in a multidimensional space of parameters.

## ACKNOWLEDGMENT

I am extremely thankful to Elias Metral, my CERN host during my FNAL-LARP long-term visit to CERN – not only for his permanently warm hospitality, but also for innumerable extremely useful discussions. I am also grateful to Stephane Fartoukh, Nicolas Mounet and Elena Shaposhnikova for a regular exchange of ideas related to a content of this paper. My special thanks are to Simon White for his help with NHT benchmarking. I appreciate a great support of Fermilab and LARP management for my long-term visit to CERN.

FNAL is operated by Fermi Research Alliance, LLC under Contract No. De-AC02-07CH11359 with the United States Department of Energy.